\documentclass[%
preprint,
 amsmath,amssymb,
 pra,
]{revtex4-1}

\usepackage{graphicx}
\usepackage{dcolumn}
\usepackage{bm}

\usepackage{color}
\usepackage{amssymb}
\usepackage{amsfonts}

\usepackage[colorlinks,urlcolor=blue,linkcolor=blue,citecolor=blue,anchorcolor=blue]{hyperref}

\begin{document}


\title{Beam splitter and combiner based on Bloch oscillations in spatially modulated waveguide arrays}

\author{Yiqi Zhang$^1$}
\email{zhangyiqi@mail.xjtu.edu.cn}
\author{Milivoj R. Beli\'c$^{2}$}
\author{Weiping Zhong$^3$}
\author{Feng Wen$^1$}
\author{Yang Guo$^1$}
\author{Yao Guo$^1$}
\author{Keqing Lu$^4$}
\author{Yanpeng Zhang$^{1}$}
\affiliation{%
 $^1$Key Laboratory for Physical Electronics and Devices of the Ministry of Education \& Shaanxi Key Lab of Information Photonic Technique,
Xi'an Jiaotong University, Xi'an 710049, China \\
$^2$Science Program, Texas A\&M University at Qatar, P.O. Box 23874 Doha, Qatar \\
$^3$Department of Electronic and Information Engineering, Shunde Polytechnic, Shunde 528300, China \\
$^4$School of Electronics and Information Engineering, Tianjin Polytechnic University, Tianjin 300387, China
}%

\date{\today}

\begin{abstract}
  \noindent
  We numerically investigate the light beam propagation in periodic waveguide arrays which are elaborately modulated with certain structures.
  We find that the light beam may split, coalesce, deflect, and be localized during propagation in these spatially modulated waveguide arrays.
  All the phenomena originate from Bloch oscillations,
  and supply possible method for fabricating on-chip beam splitters and beam combiners.
\end{abstract}

\maketitle

%
\section{Introduction}

\begin{figure}[htbp]
\centering
  \includegraphics[width=0.6\columnwidth]{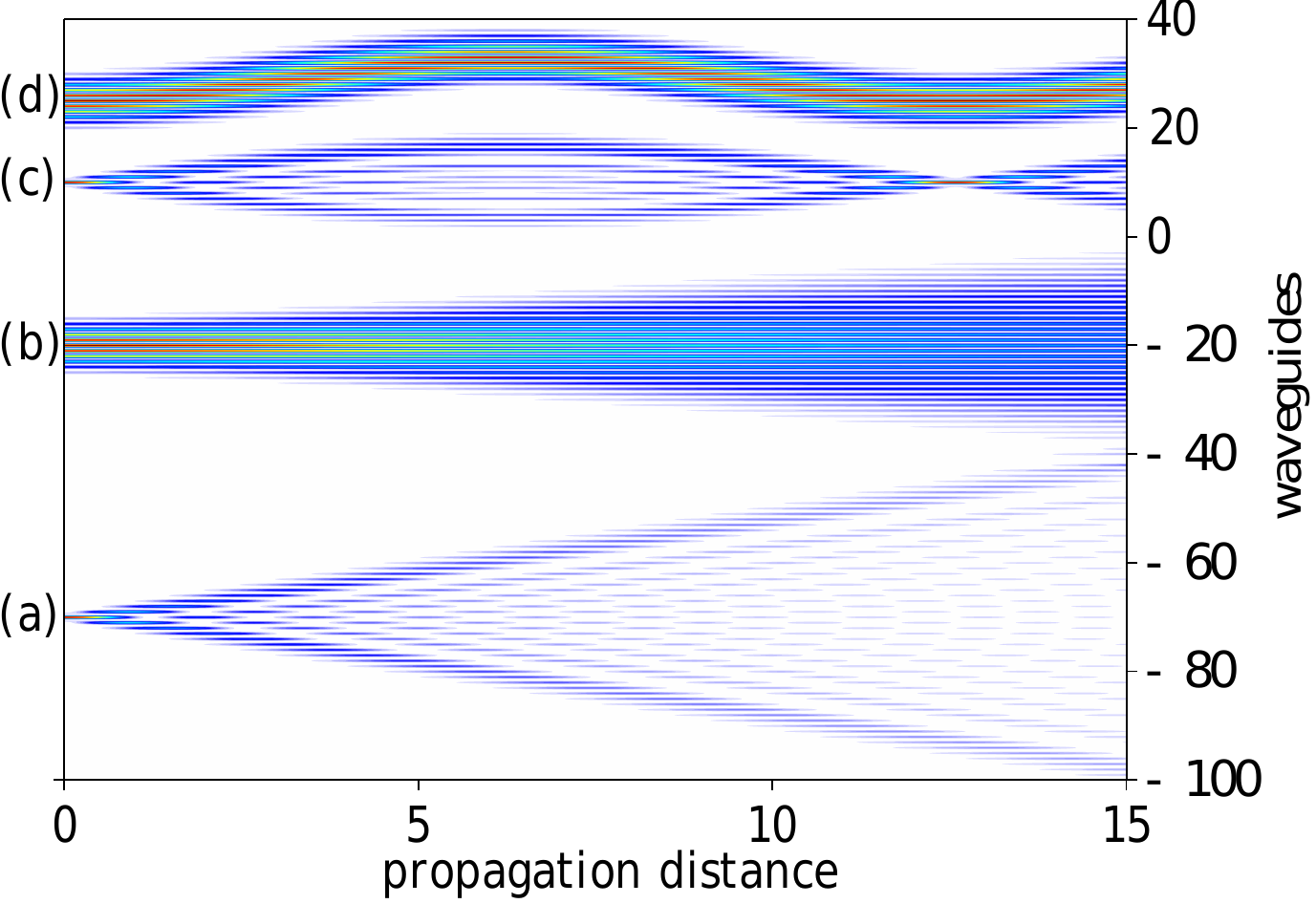}
  \caption{(Color online)
  (a) Discrete diffraction when only one waveguide is excited.
  (b) Same as (a) but with several waveguide being excited.
  (c) Bloch oscillation with one waveguide being excited.
  (d) Same as (c) but several waveguides are excited.
  In this figure, we let $\alpha=0$ for $m<0$ and $\alpha=0.5$ for $m\ge0$.}
  \label{fig1}
\end{figure}

It is well known that a light beam will undergo discrete diffraction in waveguide arrays or
modulated photonic crystals \cite{eisenberg.prl.81.3383.1998, eisenberg.prl.85.1863.2000, pertsch.prl.88.093901.2002, christodoulides.nature.424.6950.2003, lederer.pr.463.1.2008, szameit.prl.104.223903.2010, szameit.prl.106.193903.2011, garanovich.pr.518.1.2012, golshani.prl.113.123903.2014}.
Such discrete diffraction is resulted from the so-called evanescent coupling among adjacent waveguidies \cite{szameit.jpb.43.163001.2010}.
In the coupled mode regime, the evolution of light in one-dimensional periodic system can be described by
\begin{equation}\label{eq1}
  i \frac{\partial \psi_m}{\partial z} + C (\psi_{m-1} + \psi_{m+1}) = 0
\end{equation}
with $\psi_m$ being the modal amplitude in the $m$th waveguide and $C$ the coupling coefficient that we assume $C=1$.
The analytical solution of this coupled mode equation can be casted as $\psi_m(z)=J_{-m}(2z)\exp(-im\pi/2)$.
In Fig. \ref{fig1}(a), the discrete diffraction is displayed numerically by using the split-step Fourier transform method in frequency domain
when only one waveguide is excited, which agrees with the analytical solution.
While if more than one waveguides are excited, the discrete diffraction is shown as in Fig. \ref{fig1}(b).
Note that the color scale is arbitrary unit (which is also true throughout the article) because the system is linear.
As the case in continuum media, when nonlinearity is introduced, such diffraction may be balanced,
that the localized trapped beams (discrete solitons) will be obtained \cite{fleischer.nature.422.147.2003, efremidis.prl.91.213906.2003, fleischer.oe.13.1780.2005, szameit.oe.13.10552.2005, szameit.oe.14.6055.2006, garanovich.oe.15.9547.2007, szameit.pra.78.031801.2008, kartashov.rmp.83.247.2011}.

To localize a beam in bulk or atomic media, nonlinearity is required generally \cite{zhang.ol.38.4585.2013, zhang.oe.22.7160.2014, zhang.ol.37.4507.2012, zhang.cpb.18.2359.2009}.
However, to localize the beam in discrete systems, nonlinearity is not a must.
By elaborating the waveguide arrays, light beam may exhibit the so-called \textit{Bloch oscillation} phenomenon during propagation,
which prohibits the happening of discrete diffraction and localizes the beam linearly \cite{peschel.ol.23.1701.1998, pertsch.prl.83.4752.1999, morandotti.prl.83.4756.1999, usievich.os.97.790.2004, chiodo.ol.31.1651.2006, dreisow.prl.102.076802.2009}.
The basic model for investigating optical Bloch oscillation can be written as \cite{peschel.ol.23.1701.1998}
\begin{equation}\label{eq2}
  i \frac{\partial \psi_m}{\partial z} + \alpha m \psi_m + C (\psi_{m-1} + \psi_{m+1}) = 0,
\end{equation}
in which $\alpha$ is used to weigh the transverse refractive index gradient.
The corresponding analytical solution is $\psi_m(z)=J_{-m}[4\sin(\alpha z/2)/\alpha]\exp[im(\alpha z - \pi)/2]$ \cite{peschel.ol.23.1701.1998}.
In Figs. \ref{fig1}(c) and \ref{fig1}(d),
the Bloch oscillations with $\alpha=0.5$ are exhibited numerically which correspond to Figs. \ref{fig1}(a) and \ref{fig1}(b), respectively.
In addition, by designing and fabricating the periodic waveguides arrays inventively
would lead to other interesting optical phenomena \cite{keil.ol.37.809.2012, stutzer.ol.38.1488.2013, heinrich.ol.39.6130.2014, rechtsman.nature.496.196.2013}.
It is worth mentioning that such phenomena associated with periodic systems can be used to emulate the quantum events \cite{longi.lpr.3.243.2009, szameit.jpb.43.163001.2010, szameit.prl.104.150403.2010, keil.pra.81.023834.2010, keil.prl.107.103601.2011, grafe.sc.2.562.2012}.

In this article, enlightened by Bloch oscillations that happened in periodic waveguide arrays,
we design the waveguide arrays on purpose which would help us manage light behaviors during propagation
(e.g. fission, coalescence, deflection, localization, etc.).
The periodic waveguide arrays with certain structures studied in this article may have potential applications
in producing on-chip optical devices, such as beam splitters and beam combiners.

\section{Only waveguide arrays with $m\le0$ are linearly modulated}

To design the waveguides, we rewrite the parameter $\alpha$ as $\alpha_m$, which is dependent on $m$.
Therefore, Eq. (\ref{eq2}) will be rewritten as
\begin{equation}\label{eq3}
  i \frac{\partial \psi_m}{\partial z} + \alpha_m m \psi_m + C (\psi_{m-1} + \psi_{m+1}) = 0.
\end{equation}

To be frank, the coupled mode equation [Eqs. (\ref{eq1}) and (\ref{eq2})] should be written
in different forms in different regions when there is interface in the waveguide arrays \cite{szameit.njp.10.103020.2008, rechtsman.ol.36.4446.2011}.
However, we assume that the coupling constant between two adjacent waveguides is always the same,
and there is no difference of the propagation constants
in the isolated waveguides between different regions of the array in this article.
Therefore, the coupled mode equation will remain the same even though there is interface in the waveguide arrays.

\begin{figure}[htbp]
\centering
  \includegraphics[width=0.65\columnwidth]{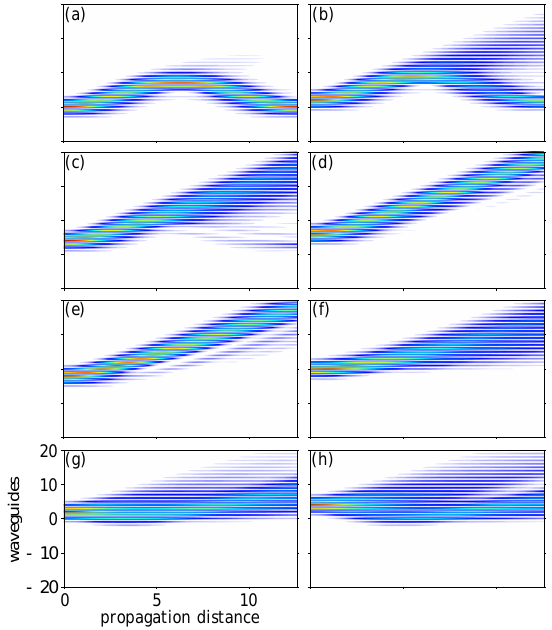}
  \caption{(Color online) Propagation of the light beam in periodic waveguide arrays
  with $\alpha=0.5$ for $m\le0$ and $\alpha=0$ for $m>0$.
  (a)-(h) Peaks of the incident beams locate at the waveguides $-10$, $-8$, $-6$, $-4$, $-2$, 0, $2$, and $4$, respectively.}
  \label{fig2}
\end{figure}

In this section, we will let $\alpha_m=0$ if $m>0$ and be a constant if $m\le0$, for example
\begin{equation}\label{eq4}
\alpha_m=\left\{
  \begin{array}{ll}
  0.5,& m\le0; \\
  0,&   m>0
  \end{array}
  \right.
\end{equation}
which is quite similar to the case shown in Fig. \ref{fig1},
but we may see Bloch oscillations in the region $m<0$ and discrete diffractions in $m>0$ if $|m|$ is large enough.
Since the Bloch oscillation will spread over $\sim 8/\alpha_m = 16$ waveguides for this case,
if the waveguide excited by incident beam peak is $m\le-12$,
a complete Bloch oscillation will be exhibited;
or, the beam energy will be coupled into the waveguides $m>0$.
Here, for convenience, we define the waveguides with $m>0$ as positive waveguides,
those with $m<0$ as negative waveguides, and that with $m=0$ as 0 waveguide, respectively.
In Figs. \ref{fig2}(a)-\ref{fig2}(h), we show the propagations of the incident beam when the waveguide excited by the beam peak is
$-10,\,-8,\,-6,\,-4,\,-2,\,0,\,2$ and 4, respectively.

In Fig. \ref{fig2}(a), the beam energy coupled into the positive waveguides is tiny, so the Bloch oscillation is nearly unaffected.
When the excited waveguide excited by the beam peak is $-8$, as shown in Fig. \ref{fig2}(b),
the beam energy coupled into the positive waveguides increases and exhibits Zener-tunneling-like behaviors \cite{longhi.prl.101.193902.2008, dreisow.prl.102.076802.2009}, which is due to the discrete diffraction of part of the oscillated beam that coupled into the positive waveguides.
When the excited waveguide is closer to the 0 waveguide, as displayed in Figs. \ref{fig2}(c)-\ref{fig2}(e),
more of the energy [Fig. \ref{fig2}(c)] and even all the energy [Figs. \ref{fig2}(d) and \ref{fig2}(e)] is coupled into the positive waveguides.
Especially in Fig. \ref{fig2}(d), the beam seems to be localized to some extent.
The reason for this phenomenon is that the beam turns to and at the same time the energy is coupled into the positive waveguides,
thus the diffraction is equivalent to that of an obliquely incident beam \cite{pertsch.prl.88.093901.2002}.

If the waveguide excited by the beam peak is right the 0 waveguide, as shown in Fig. \ref{fig2}(f),
the incident beam energy distributes in the positive and negative waveguides equally,
which will undergo discrete diffraction and Bloch oscillation, respectively.
On one hand, the discrete diffraction will be reflected into the positive waveguides when it reaches the 0 waveguide;
on the other hand, Bloch oscillation will couple the energy from negative waveguides into positive waveguides.
As a result, the beam mainly exhibits discrete diffraction in the positive waveguides.

When the waveguide excited by the beam peak is positive,
most of the beam energy will undergo discrete diffraction.
And the diffraction will be totally internally reflected into the positive waveguides when it approaches to the 0 waveguide,
as shown in Figs. \ref{fig2}(g) and \ref{fig2}(h).
Besides, there is Goos-H\"anchen shift \cite{rechtsman.ol.36.4446.2011} clearly in the negative waveguides.

\section{Beam splitters based on V-type modulated waveguide arrays}

Instead of only modulating the negative waveguides,
we will modulate both the positive and negative waveguides by defining $\alpha_m$ as
\begin{equation}\label{eq5}
\alpha_m=\left\{
  \begin{array}{lll}
  -0.5,& m\le0; \\
  \:\:\;0.5,&   m>0.
  \end{array}
  \right.
\end{equation}
Therefore, the modulated waveguides will be exhibit V-type,
which means the bigger $|m|$ the higher the index ramp.
Thus, two beams respectively launched into the negative and positive waveguides will be first ``repelled'' and then ``attracted'' to oscillate
along the opposite directions.
That's to say, the beams will be away from the 0 waveguide within the propagation distance $z<\pi/|\alpha_m|$ and
close to the 0 waveguide within $\pi/|\alpha_m| \le z<2\pi/|\alpha_m|$.
They do not affect each other if they do not overlap with each other at the initial place
(i.e., at least one of them is not across the 0 waveguide).
While if the incident beam crosses the 0 waveguide, it will be divided.
From this point of view, this would be used to make beam splitters \cite{breid.njp.8.110.2006}.

\begin{figure}[htbp]
\centering
  \includegraphics[width=0.6\columnwidth]{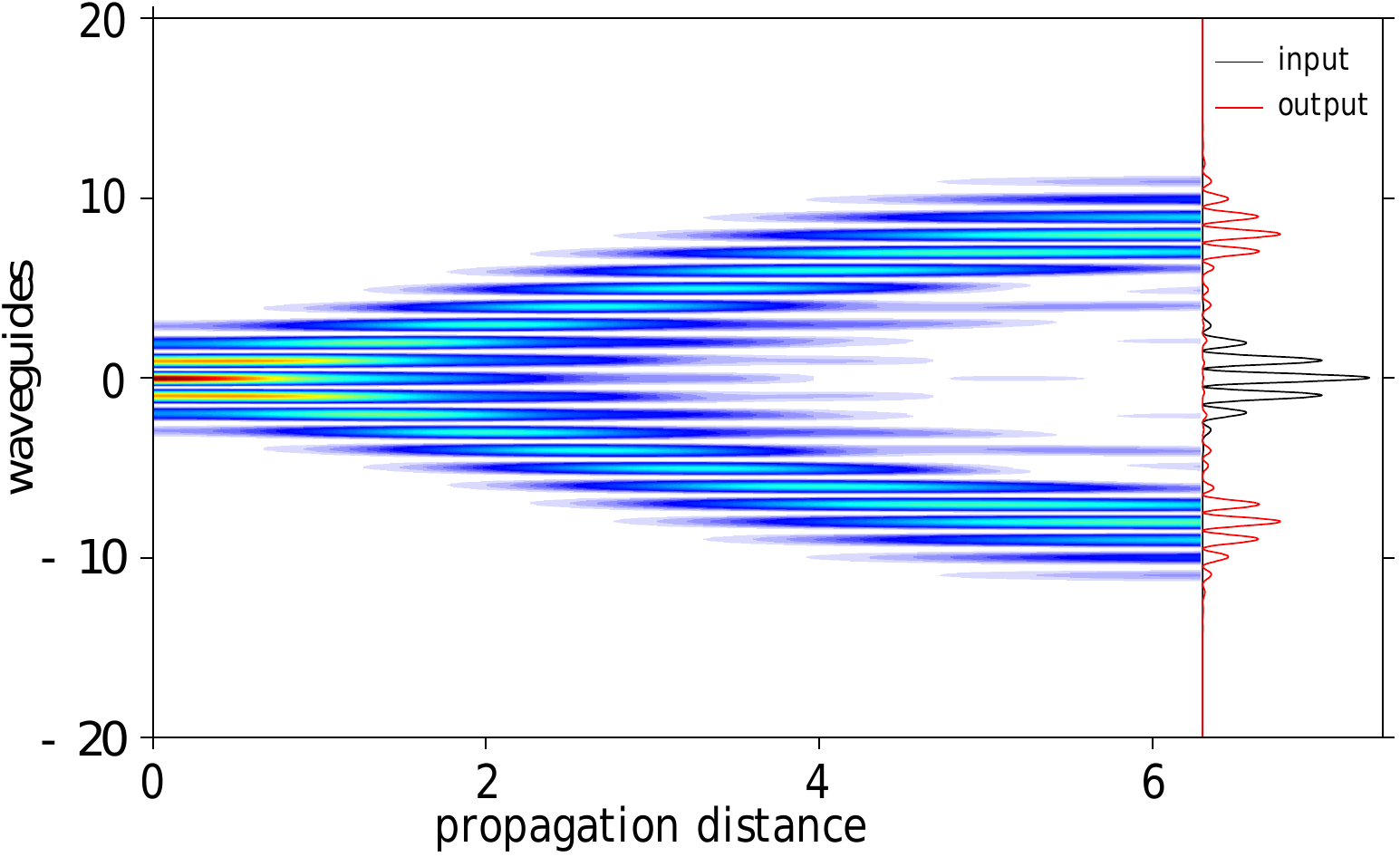}
  \caption{(Color online) A Gaussian beam splits during propagation in a V-type modulated waveguide arrays.
  The 0 waveguide is excited by the beam peak.
  Curves in the right panel are the corresponding beam intensities at input and output places, respectively.}
  \label{fig3}
\end{figure}

In Fig. \ref{fig3}, we show the propagation of a Gaussian beam in the V-type modulated waveguide arrays within the half period of Bloch oscillation,
and the 0 waveguide is excited by the beam peak.
Clearly, the beam is equally divided into the positive and negative waveguides during propagation.
In addition, as shown by the input and output beam intensities in the right panel in Fig. \ref{fig3},
the maximum intensities of the out beam locate at the $\pm8$ waveguides,
and the divided humps can be recognized from the initial hump apparently.
In a word, both the efficiency and resolution of the fission are quite high,
which means that such fission can be used to produce beam splitters.
We would like to emphasize that this kind of beam splitters are only based on Bloch oscillations,
which is different from the situation reported previously \cite{breid.njp.8.110.2006}.
We believe the beam splitter is much easier to be implemented and
can be conveniently tuned (such as fission width and fission length) through the parameter $\alpha_m$.

\section{Beam combiners based on $\Lambda$-type modulated waveguide arrays}

It is interesting to wonder what will happen if the modulation is $\Lambda$-type -- an inverted case of V-type.
For this case, $\alpha_m$ will be written as
\begin{equation}
\alpha_m=\left\{
  \begin{array}{lll}
  \:\:\;0.5,& m\le0; \\
     -0.5,&   m>0.
  \end{array}
  \right.
\end{equation}
Same as the case in V-type modulated waveguide arrays,
the incident beam will exhibit Bloch oscillation if the excited waveguide number $|m|$ is not smaller than $\sim 6/|\alpha_m|$.

\begin{figure*}[htbp]
\centering
  \includegraphics[width=0.6\columnwidth]{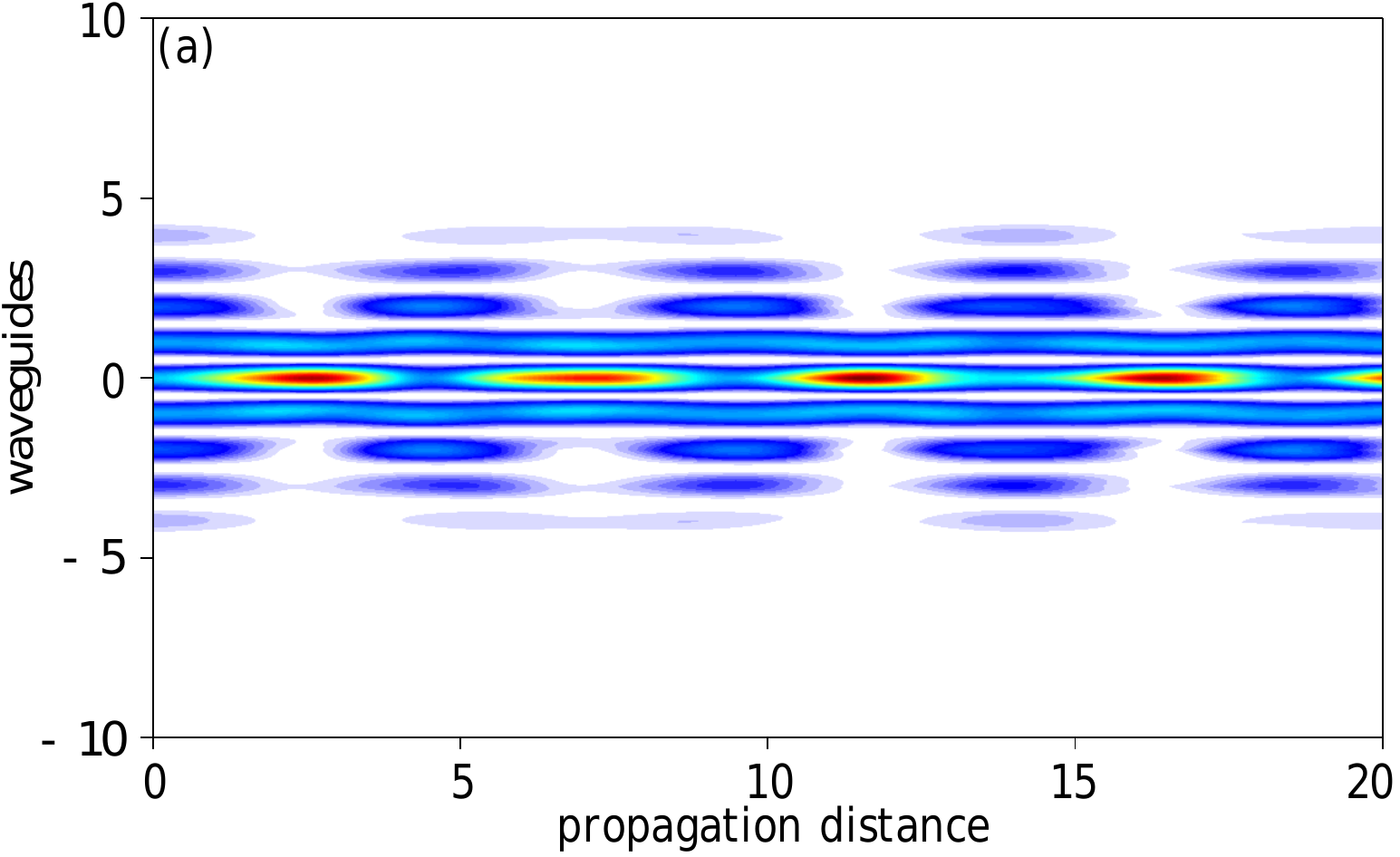}
  \includegraphics[width=0.6\columnwidth]{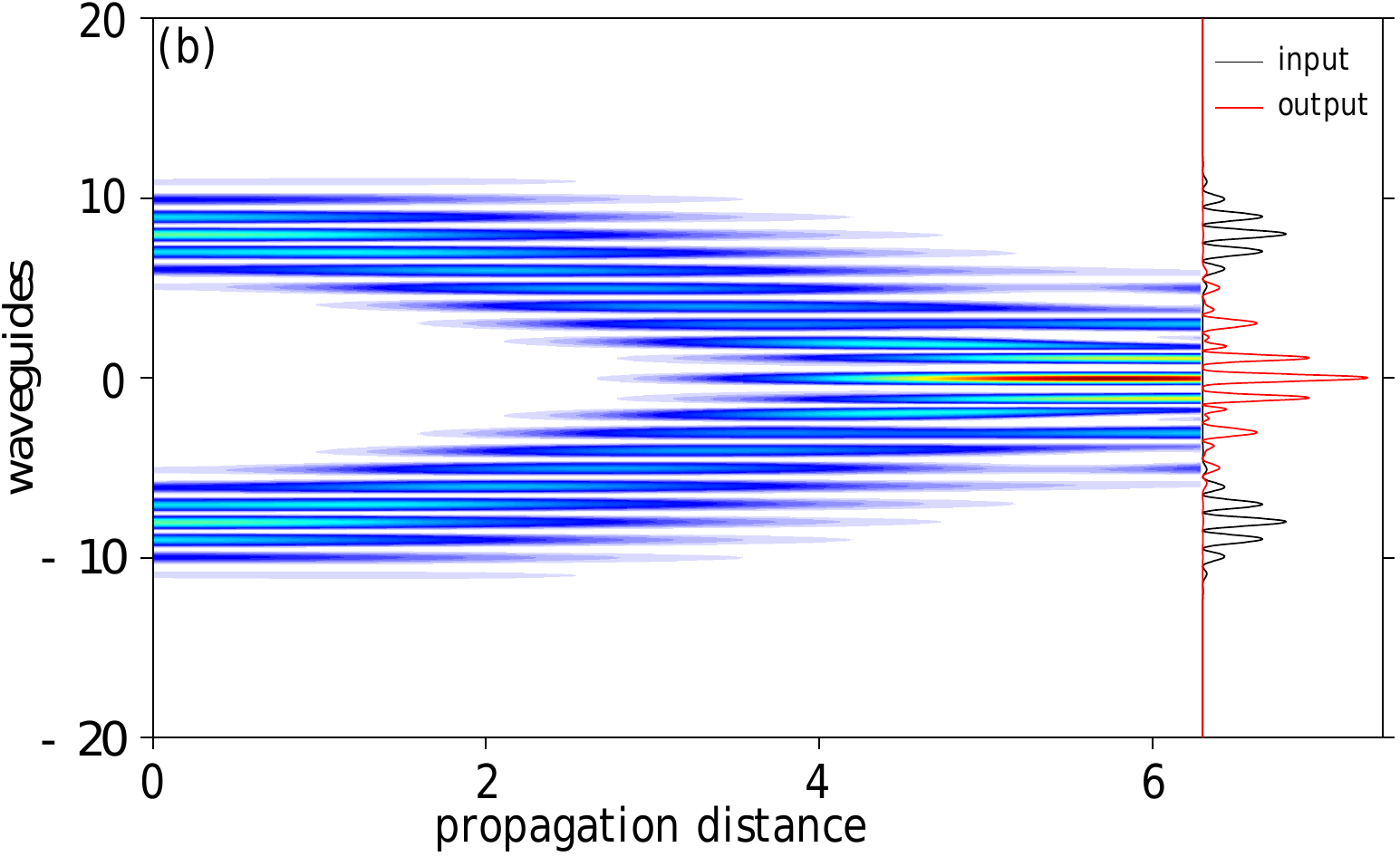}
  \caption{(Color online)
  (a) Propagation of a Gaussian beam in the $\Lambda$-type modulated waveguide arrays with the 0 waveguide excited by the beam peak.
  A Fermi-Pasta-Ulam recurrence-like localization happens during propagation.
  (b) Two Gaussian beams respectively launched from positive and negative waveguides coalesce during propagation.
  Curves in the right panel are the corresponding beam intensities at input and output places, respectively.}
  \label{fig4}
\end{figure*}

We first investigate the propagation of a Gaussian beam with the peak exciting the 0 waveguide, as displayed in Fig. \ref{fig4}(a).
From Fig. \ref{fig4}(a), we can see that the beam energy is first attracted to the 0 waveguide,
then repelled into the positive and negative waveguides, and this process happens periodically,
which is quite similar to the Fermi-Pasta-Ulam recurrence discovered in fiber \cite{simaeys.prl.87.033902.2001, mussot.prx.4.011054.2014}.
The reason for this phenomenon is quite trivial,
because this is Bloch oscillation in essence,
which is a result of combining effect of discrete diffraction, totally internal and Bragg reflections \cite{peschel.ol.23.1701.1998}.
It is worth mentioning that such linear localization spreads over waveguide arrays no more than the incident beam does
if more than 1 waveguide is excited by the beam and the 0 waveguide is excited by the peak,
which is independent on the parameter $\alpha_m$.

If we launch two beams into the positive and negative waveguides,
they will first ``attract'' with each other when they begin to oscillate, which is different from the V-type case.
Therefore, if the excited waveguide number $|m| \le 6/|\alpha_m|=12$, they will coalesce during propagation.
In Fig. \ref{fig4}(b), we show this kind of coalescence.
The two Gaussian beams which respectively excite the $\pm8$ waveguides, oscillate to the 0 waveguide.
Since $|m|<12$, both the two beams can reach the 0 waveguide, which cause the two beams coalesce.
Similar to Fig. \ref{fig3}, we also display the input and output intensities in Fig. \ref{fig4}(b).
It is no doubt that the coalesced beam is Gaussian-like,
even though the intensities distributed in the $\pm3$ waveguides are refrained
due to destructive interference between the oscillated beam and its corresponding reflection.
From Fig. \ref{fig4}(b), we can see that such kind of coalescence can be used to produce beam combiners,
which is also just from the Bloch oscillations in waveguide arrays.
Same as the beam splitter shown in Fig. \ref{fig3},
such beam combiner possesses high  work efficiency and resolution,
and meanwhile it is conveniently to be tuned.

\section{conclusion}

In conclusion, we investigated propagation of beams in linearly modulated waveguide arrays.
By elaborately design the modulation forms,
beams can split, coalesce, deflect, and be linearly localized during propagation,
which can be understood according to Bloch oscillations.
We also find that the results can be conveniently tuned through the parameter $\alpha_m$.
Based on the beam fission and coalescence,
we believe that prototype optical devices, such as beam splitters and beam combiners,
can be simply produced with high work efficiency and resolution.

\section*{Acknowledgement}

This work was supported by the 973 Program (2012CB921804),
KSTIT of Shaanxi province (2014KCT-10),
NSFC (61308015, 11104214, 61108017, 11104216, 61205112),
CPSF (2014T70923, 2012M521773),
and the Qatar National Research Fund NPRP 6-021-1-005 project.

\bibliographystyle{osajnl}
\bibliography{refs}

\end{document}